\documentclass[default,iicol,sn-mathphys]{sn-jnl}% Default with double column layout
\usepackage{subfigure}

\usepackage{float}
 \usepackage{etoolbox}
 \pdfoutput=1
\makeatletter
\patchcmd{\ps@headings}
{\hbox to \hsize{\hfill Springer Nature 2021 \LaTeX\ template\hfill}}
{\hbox to \hsize{}}
{}
{}
\patchcmd{\ps@titlepage}
{\hbox to \hsize{\hfill Springer Nature 2021 \LaTeX\ template\hfill}}
{\hbox to \hsize{}}
{}
{}
\makeatother
\makeatletter
\patchcmd{\ps@headings}
{\hbox to \hsize{\hfill Springer Nature 2021 \LaTeX\ template\hfill}}
{\hbox to \hsize{}}
{}
{}
\patchcmd{\ps@headings}
{\hbox to \hsize{\hfill Springer Nature 2021 \LaTeX\ template\hfill}}
{\hbox to \hsize{}}
{}
{}
\patchcmd{\ps@titlepage}
{\hbox to \hsize{\hfill Springer Nature 2021 \LaTeX\ template\hfill}}
{\hbox to \hsize{}}
{}
{}
\makeatother

\jyear{2021}%

\theoremstyle{thmstyleone}%
\theoremstyle{thmstyletwo}%

\theoremstyle{thmstylethree}%
\usepackage{arydshln}
\usepackage{booktabs}
\usepackage{color}
\usepackage{array}
\usepackage{tabularx}
\usepackage{adjustbox}
\usepackage{mathrsfs}
\usepackage{mathtools}
\usepackage{extarrows}
\usepackage{multirow}
\usepackage{graphicx}
\usepackage{amssymb}
\usepackage{bbding}
\usepackage{overpic}
\usepackage{soul}
\usepackage[marginal]{footmisc}
\usepackage[misc]{ifsym} 

\begin{document}

\title[Article Title]{LKFormer: Large Kernel Transformer for Infrared Image Super-Resolution}

\author{\fnm{Feiwei} \sur{Qin}$^{1}$, \fnm{Kang} \sur{Yan}$^{1}$, \fnm{Changmiao} \sur{Wang}$^{2}$, \fnm{Ruiquan} \sur{ Ge}$^{1,*}$, \fnm{Yong} \sur{Peng}$^{1}$, \\ \fnm{Kai} \sur{Zhang}$^{3,*}$}

% \author{\fnm{Yong} \sur{Peng}$^{1}$,  \fnm{Kai} \sur{Zhang}$^{3,*}$}
% \author{\fnm{Kang} \sur{Yan}$^{1}$}

% \author{\fnm{Changmiao} \sur{Wang}$^{2}$}
% \author{\fnm{Ruiquan} \sur{ Ge}$^{1,*}$}
% \author{\fnm{Yong} \sur{Peng}$^{1}$}
% \author*{\fnm{Kai} \sur{Zhang}$^{3,*}$}
% \address[author1]{School of Computer Science and Technology, Hangzhou Dianzi University, China}

% \address[author4]{Shenzhen Research Institute of Big Data, China}

% \address[author5]{CVL, ETH Zurich, Switzerland}

% \affil[1]{ \orgaddress{\street{School of Computer Science and Technology, Hangzhou Dianzi University}, \country{China}}}

% \affil[2]{\orgaddress{\street{Shenzhen Research Institute of Big Data},  \country{China}}}

% \affil[3]{\orgaddress{\street{CVL, ETH Zurich},  \country{Switzerland}}}

% \affil[3]{\orgdiv{Department}, \orgname{Organization}, \orgaddress{\street{Street}, \city{City}, \postcode{610101}, \state{State}, \country{Country}}}


\abstract{Given the broad application of infrared technology across diverse fields, there is an increasing emphasis on investigating super-resolution techniques for infrared images within the realm of deep learning. Despite the impressive results of current Transformer-based methods in image super-resolution tasks, their reliance on the self-attentive mechanism intrinsic to the Transformer architecture results in images being treated as one-dimensional sequences, thereby neglecting their inherent two-dimensional structure. Moreover, infrared images exhibit a uniform pixel distribution and a limited gradient range, posing challenges for the model to capture effective feature information. Consequently, we suggest a potent Transformer model, termed Large Kernel Transformer (LKFormer), to address this issue. Specifically, we have designed a Large Kernel Residual Attention (LKRA) module with linear complexity. This mainly employs depth-wise convolution with large kernels to execute non-local feature modeling, thereby substituting the standard self-attentive layer. Additionally, we have devised a novel feed-forward network structure called Gated-Pixel Feed-Forward Network (GPFN) to augment the LKFormer's capacity to manage the information flow within the network. Comprehensive experimental results reveal that our method surpasses the most advanced techniques available, using fewer parameters and yielding considerably superior performance.}

%%================================%%
%% Sample for structured abstract %%
%%================================%%

% \abstract{\textbf{Purpose:} 
% 
% \textbf{Conclusion:} The abstract serves both as a general introduction to the topic and as a brief, non-technical summary of the main results and their implications. The abstract must not include subheadings (unless expressly permitted in the journal's Instructions to Authors), equations or citations. As a guide the abstract should not exceed 200 words. Most journals do not set a hard limit however authors are advised to check the author instructions for the journal they are submitting to.}


\maketitle
\renewcommand{\thefootnote}{}
\footnote[1]{\textrm{\Letter} \fnm{Ruiquan} \sur{ Ge}}

\footnotetext{gespring@hdu.edu.cn}

\footnote[1]{\textrm{\Letter} \fnm{Kai} \sur{Zhang}}
\renewcommand{\thefootnote}{}
\footnotetext{kai.zhang@vision.ee.ethz.ch}

\renewcommand{\thefootnote}{}
\footnote[1]{$^{1}$\orgaddress{\street{School of Computer Science and Technology, Hangzhou Dianzi University}, \country{China}}}
\footnote[2]{$^{2}$\orgaddress{\street{Shenzhen Research Institute of Big Data},  \country{China}}}
\footnote[3]{$^{3}$\orgaddress{\street{CVL, ETH Zurich},  \country{Switzerland}}}

\clearpage
\section{Introduction}\label{sec1}
Image Super-Resolution (SR) is a prevalent low-level vision issue that has risen to prominence in recent years. The goal of image SR is to reconstruct high-quality images from their degraded, low-resolution counterparts by eliminating degradation artifacts. 
While numerous learning-based methods have been previously developed for visible image SR, there is a burgeoning need to improve the resolution of infrared images. Infrared images can offer invaluable information in scenarios where visibility is compromised due to severe weather or challenging environments. They are extensively deployed in sectors such as security, medicine, construction, energy, and environmental studies \cite{sousa2017review,lopez2017application,kirimtat2018review}. Despite the great potential of infrared images, creating high-quality infrared image datasets through image SR networks is a challenge due to low contrast and poor perceptual quality in low-resolution infrared images.

Convolutional Neural Networks (CNNs) have become a fundamental component in image SR models and have showcased significant reconstruction success. However, these CNN-based methods \cite{lim2017enhanced,wang2018esrgan,zhang2018image,zhang2021plug} are confined to local structural information due to the limited receptive field of convolutional operators. This limitation makes them less capable of addressing long-range degenerate patterns present in low-resolution infrared images. To surmount these limitations, researchers have begun to explore the Transformer structure \cite{vaswani2017attention},a prevalent model in natural language processing. With the notable work of SwinIR \cite{liang2021swinir}, which incorporates the self-attentive (SA) mechanism, the Transformer structure has demonstrated remarkable success in single-image SR tasks. Unlike CNNs, SA mechanisms model non-local information, rendering them ideal for high-quality image restoration. However, while these mechanisms achieve superior results, the quadratic growth of computational complexity in relation to spatial resolution restricts their application to infrared terminal devices with limited computational resources. Moreover, SA mechanisms neglect the adaptation of channel dimensions and prohibit modeling of the local invariant property harnessed by CNNs.

To maximize the powerful feature extraction capabilities of CNNs and achieve global information modeling, we propose an innovative and efficient Transformer model, termed the Large Kernel Transformer (LKFormer). This model capitalizes on a large receptive field to foster increased feature interactions, thereby overcoming the limitations of previous CNN-based SR methods, which struggled to model long distances. Specifically, we introduce a Large Kernel Residual Attention (LKRA) block with linear complexity to supplant the standard multi-head self-attention. The LKRA block encodes local structural information and long-range pixel dependencies by combining depth-wise convolution with variable kernel size, while also utilizing $1 \times 1$ convolution to facilitate information interaction between channels. Furthermore, given the absence of prior information in low-resolution infrared images, it is crucial to fully leverage the extracted features during SR while also exploring their underlying relationships. For this purpose, the residual connection is incorporated multiple times within the LKRA block.

In the Transformer architecture, the feed-forward network (FN) operates as a critical component of the model. The standard FN consists of two fully connected layers that encompass a non-linear layer, facilitating local non-linear transformations of features. This study introduces an advanced FN design termed 'gated-pixel feed-forward network' (GPFN), specifically engineered to enhance the Transformer structure for the intricate pixel prediction task associated with image SR. This enhancement is achieved by fine-tuning the original FN through the incorporation of a pixel attention branch subsequent to the non-linear layer. This extra branch generates an attention map, denoted as $\mathcal{M}$, possessing identical dimensions to the current feature $\mathcal{F}$, followed by executing a dot product operation on $\mathcal{F}$ and $\mathcal{M}$. This methodology effectively amplifies the weight of pertinent content, thereby enabling succeeding network layers to concentrate more keenly on the finer characteristics of the image. Collectively, these modifications contribute to superior performance and high-quality output.

The main contributions of this work can be summarized as follows:
\begin{itemize}
\item[$\bullet$] We introduce a unique Transformer model, termed as LKFormer, aimed at achieving explicit modeling of both local and global range dependencies in infrared images.
\item[$\bullet$] We propose an inventive module named LKRA that enables the aggregation of both local and non-local pixel interactions. This is primarily accomplished by integrating separable residual depth-wise convolutional blocks with varying receptive fields.
\item[$\bullet$] We present a reconfigured and highly effective feedforward network, founded on pixel attention. This network is capable of further adjusting the weights of pertinent features to more efficiently facilitate the SR reconstruction of infrared images.

\item[$\bullet$] Extensive experimental results demonstrate that the LKFormer outperforms existing state-of-the-art methods in terms of performance.

\end{itemize}

\section{Related Work}\label{sec2}

\subsection{Deep Learning-based Single-image SR}\label{subsec1}
\textbf{CNN-based methods.} Owing to the robust representational capabilities of CNNs, and their faculty to utilize the feature hierarchy in learning mapping relationships between input and target images, several image SR models \cite{dong2015image,kim2016deeply,zhang2017learning} were based on CNN architectures during the initial rise of deep learning. To harness the capacity of CNN architectures to represent abstract and high-level semantic information, some methodologies \cite{kim2016accurate,cavigelli2017cas} began experimenting with residual connectivity. This approach allowed CNNs to capture more information while retaining efficient training speed and accuracy. Furthermore, some techniques \cite{zhang2018residual,zhang2020residual} employed dense connectivity to aggregate hierarchical features for superior image reconstruction. As the exploration of attention mechanisms in CNN architectures gained traction, an increasing number of image SR methods \cite{dai2019second,niu2020single,zhao2020efficient} began incorporating either channel attention or spatial attention modules, thus allowing selective focus on relevant information. However, since most of these previous methods are based on small kernel (e.g., $3\times3$) CNNs, they encounter challenges in modeling degraded distributions over extensive distances. To address this, Mei et al. \cite{52mei2021image} proposed a novel convolutional neural network structure termed non-local sparse attention. This structure is capable of globally capturing texture information by introducing non-local blocks and a sparse attention mechanism, despite the significant computation resources it requires.

\textbf{Transformer-based methods.} The Transformer model, originally developed for sequential data processing in natural language tasks, has been successfully extended to vision tasks due to its impressive performance in high-dimensional tasks. Notable examples of this extension include the Vision Transformer (ViT) \cite{dosovitskiy2020image} and Swin Transformer \cite{liu2021swin}, which have been recently utilized in image SR. These methods \cite{liang2021swinir, fang2022hybrid, chen2023activating} have exceeded the performance of previous CNN-based models, thanks to the self-attention mechanism within the Transformer structure that captures non-local features. However, it should be noted that the self-attention mechanism was initially designed to extract relationships between sequential data in natural language processing tasks. When applied to images, the mechanism treats them as 1D sequences, failing to consider their inherent 2D structure. Furthermore, while the Swin Transformer's window-based self-attention offers greater computational efficiency than global self-attention, it still presents a significant computational and memory burden. Although Zamir et al. \cite{zamir2022restormer} employ channel attention to encode global context information for reducing computational consumption, their method overlooks the significance of local features in image restoration.

\textbf{SR methods for infrared images.} Infrared imagery, facilitated by technological advancements, has become indispensable in various fields such as aircraft maintenance and object measurement \cite{si2023tri, tang2023tccfusion, ge2020occluded}. However, due to high hardware costs and complex environmental factors, the acquired infrared images often exhibit low resolution. As a result, the reconstruction of low-resolution infrared images and the generation of high-quality equivalents have emerged as paramount research focuses. Traditionally, most infrared image SR methods relied predominantly on the frequency domain or dictionary-based algorithms \cite{wang2009analysis}. Recently, however, the potent feature extraction capabilities of CNN have accelerated the development of deep learning-based infrared image SR algorithms. For instance, He et al. \citep{he2018cascaded} suggested a novel cascaded depth network approach that utilizes multiple receptive fields to enhance the SR of infrared images. In contrast to visible images, infrared images often exhibit blurred contours and edges due to variations in imaging systems. To address these challenges, Zou et al. \citep{zou2021super} designed a U-net-based model that employs a residual network to capture high-frequency and low-frequency information in infrared images. Drawing inspiration from the success of Generative Adversarial Networks (GANs) in SR tasks for visible images, scientists have leveraged GAN models to enhance the resolution of infrared images. A case in point is Huang et al. \citep{huang2021infrared}, who introduced a highly effective model, PSRGAN, that incorporates migration learning and knowledge refinement to achieve superior SR performance for infrared images using fewer parameters. Furthermore, they proposed HetSRWGAN \citep{57huang2021infrared}, which enhances training stability by integrating a new loss function. In an effort to address the issues of low resolution and loss of thermal information in infrared images, Wu et al. \citep{wu2022meta} recommended a lightweight SR method based on Meta-Transfer Learning to achieve high-resolution reconstruction of infrared images.

% \begin{figure*}[ht]%
% \centering
% \includegraphics[width=0.9\textwidth]{LKFormer-1_cropped.pdf}
% \caption{The architecture of the proposed LKFormer for lightweight infrared image SR. Here, LISHB denotes lightweight information split hybrid block. $\copyright$ and $\oplus$ represent the concatenate and sum operations, respectively}\label{fig1:LKFormer}
% \end{figure*}

% \begin{figure*}[ht]%
% \centering
% \includegraphics[width=0.9\textwidth]{LISHB-1-croped.pdf}
% \caption{The architecture of the lightweight information split hybrid block (LISHB) for deep feature extraction. }\label{fig2:LISHB}
% \end{figure*}

\subsection{Convolution for no-local range modelling}\label{subsec2.1}
AlexNet \cite{krizhevsky2017imagenet} is the classical CNN that gained significant attention in the field. It represents an early model that employed convolutional operations with large kernels. Another classical approach that leverages large kernel convolution for enhancing model performance is the Global Convolutional Network \cite{peng2017large}. The method utilizes large, symmetric, and separable convolutional kernels to improve tasks like semantic segmentation, which involve pixel-wise predictions. This allows for the computation of weights across the entire feature map. Large convolutional kernels were previously less preferred due to their high computational cost and the extensive parameter space they require. However, there has been a recent resurgence of interest in them owing to advancements in efficient convolutional techniques and the application of Transformers in vision tasks. For example, convMixer \cite{asher2022patches} utilizes depth-wise convolution with large convolutional kernels to replace the mixer component in ViT \cite{dosovitskiy2020image} or MLP \cite{tolstikhin2021mlp}. Conversely, ConvNeXt \cite{liu2022convnet} reimagines a robust base network model, inspired by ResNet \cite{he2016deep}, through $7\times7$ depth-wise convolution and other optimization strategies, yielding results comparable to or even surpassing those achieved by Swin Transformer \cite{liang2021swinir}. Acknowledging the significance of receptive fields for downstream tasks, RepLKNet \cite{ding2022scaling} develops a pure convolutional model that expands the convolutional kernel size to a record-breaking $31\times31$. In a bid to push the bounds of convolutional kernel size, SLaK \cite{liu2022more} incorporates an unmatched $51\times51$ convolution. Unlike RepLKNet, SLaK employs two complementary non-square kernels ($M \times N$ and $N \times M$) when decomposing a large kernel, thereby enhancing the training stability and memory scalability of large convolutional kernels. In this study, we adopt a similar approach to decompose large kernel convolution, thus expanding the receptive field whilst avoiding excessive computational consumption.

\section{Method}\label{sec3}
In this section, we first provide an overview of our network structure's overall pipeline. Subsequently, we delve into the specifics of the proposed Transformer layer, which forms the cornerstone of our approach. This layer primarily comprises two critical components: the LKRA and the GPFN.
% residual depth-wise convolutional attention (LKRDA) block and gated-pixel feed-forward network (GPFN).

\begin{figure*}[ht]%
\centering
\includegraphics[width=0.95\textwidth]{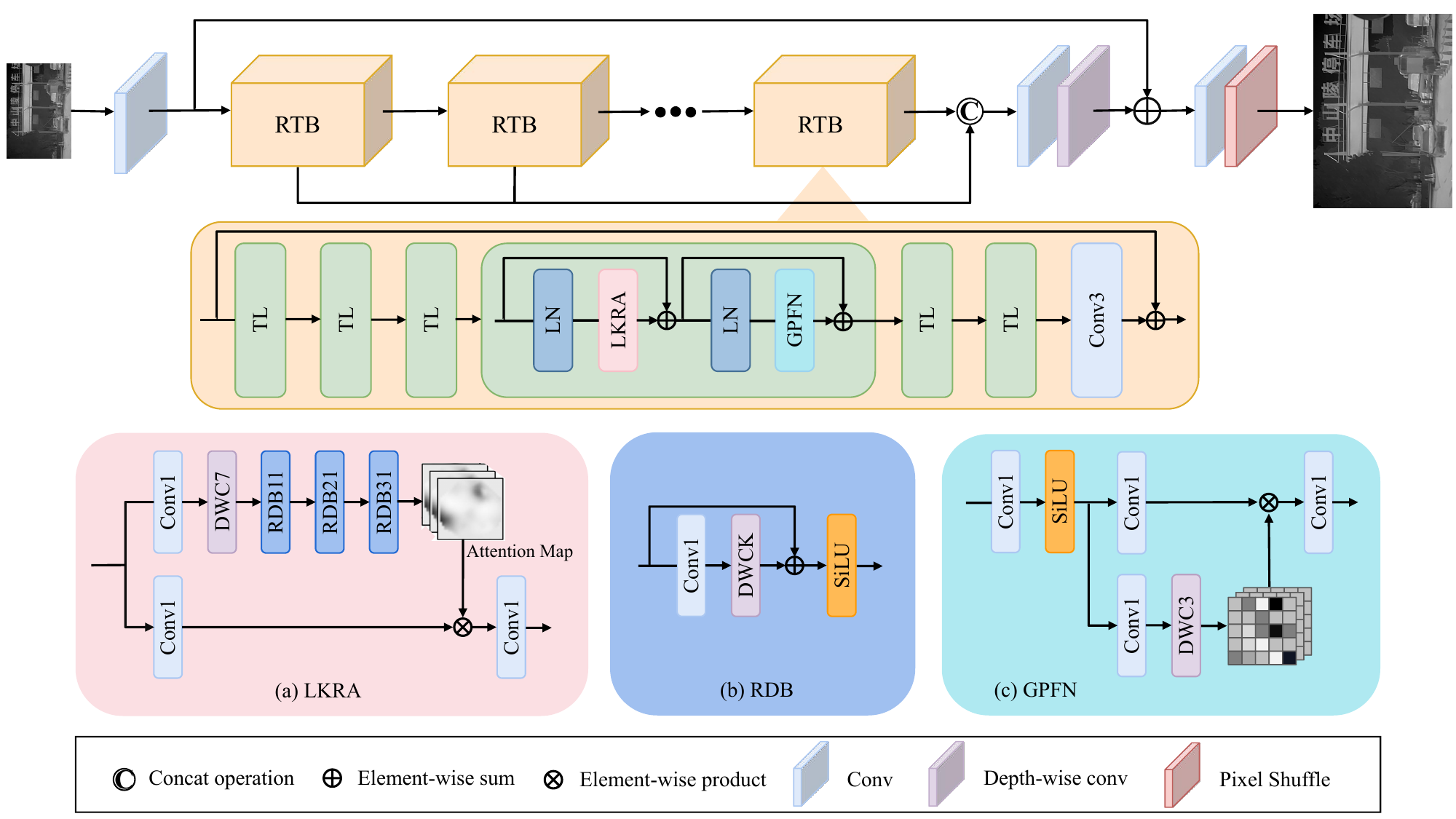}
\caption{The architecture of the proposed LKFormer for infrared image SR. Here, RDB, RTB, and TL denote the residual depth-wise convolution block, residual Transformer block, and Transformer layer respectively. LKRA stands for large kernel residual attention block, which utilizes multiple convolutional layers of different sizes to extract features and then generates an attention map as a way to achieve encoding both local structural information and long-range dependencies of the input features. GPFN refers to a module called gated-pixel feed-forward network, which adds a branch of pixel attention to the classical feed-forward neural network as a way to control the forward flow of features. Additionally, LN and DWC3 represent the layer normalization and the $3\times3$ depth-wise convolution.}\label{fig1:LKFormer}
\end{figure*}

\subsection{ Overall pipeline }\label{subsec2}
Fig.~\ref{fig1:LKFormer} illustrates the overall network structure. The network receives a low-resolution infrared image as an input and outputs a high-quality infrared image after processing. More specifically, the network consists of three main parts: First, a $3\times3$ convolutional layer performs shallow feature extraction, converting the low-resolution infrared image into a feature map. The deep feature extraction module then applies \textit{N} residual Transformer blocks (RTB) to extract spatially varying reconstructed target information distribution, while each RTB is composed by stacking multiple Transformer layers (TL) and a final convolutional layer. Jump connections are introduced in the RTB and the entire feature representation learning part in order to aid in the recovery process and stabilize training. Furthermore, to mine different levels of latent feature information and to achieve fine structural and textural detail in the reconstructed infrared images, the output of each RTB is concatenated prior to being input to the image reconstruction module. Finally, the rich features computed by the previous operations are used to estimate the recovered image in the image reconstruction module.

\begin{table*}[ht]
\renewcommand{\arraystretch}{1.5}
\begin{center}
\begin{minipage}{\textwidth}
\caption{Quantitative comparison (average PSNR/SSIM) with other compared methods for scale factor 2 and 4 on the public datasets IR700, results-A, and ESPOL FIR.}\label{tabResult}
\begin{tabular*}{\textwidth}{@{\extracolsep{\fill}}lcccccc@{\extracolsep{\fill}}}
\toprule%
%  \multirow{2}*{Method} & \multirow{2}*{Scale} & \multirow{2}*{Params (M)} & \multicolumn{2}{@{}c@{}}{IR700}  \\\cmidrule{4-5}%
% ~ & ~ & ~ & PSNR (dB) & SSIM \multicolumn{2}{c}{\multirow{2}{*}{Method}}  \\

  \multirow{2}{*}{Method} & \multirow{2}*{Scale} & \multirow{2}*{Params (M)} &\multirow{2}*{FLOPs (G)}& IR700 & results-A & ESPOL FIR \\\cmidrule{5-7}%
  ~ & ~& ~ & ~ & PSNR / SSIM & PSNR / SSIM & PSNR / SSIM  \\

\midrule
 EDSR\cite{lim2017enhanced}  & \multirow{8}{*}{$\times2$} & 40.73& 166.84  & 39.35 / 0.9481 & 37.81 / 0.9346 & 43.17 / 0.9643\\
 RRDB\cite{wang2018esrgan} &  & 26.78&   110  & 39.40  / 0.9479 & 37.83 / 0.9349 & 43.18 / 0.9649\\
 PSRGAN\cite{huang2021infrared} &  & 0.31&  2.18   & 39.13  / 0.9307 & 37.29 / 0.9212 & 42.84 / 0.9572 \\
 NLSN\cite{52mei2021image} &  & {41.80}&  171.20   & 39.41  / 0.9485 & 37.85 / 0.9353 & 43.22 / 0.9647\\
 SwinIR\cite{liang2021swinir} &  & 11.75&   51.33  & {39.49}  / {0.9489} &  37.85 / 0.9351 & 43.23 / 0.9650\\
 HAT\cite{chen2023activating}  &  & 9.47&  52.39   & 39.46  / 0.9486 & 37.90 / 0.9353  & 43.11 / 0.9645\\
 \textbf{LKFormer} &  & {6.38}& 25.81   & \textbf{39.70}  / \textbf{0.9505} & \textbf{37.91} / \textbf{0.9355} & \textbf{43.28} / \textbf{0.9655} \\
% ########################################
\midrule
EDSR\cite{lim2017enhanced}  & \multirow{8}{*}{$\times4$} & 43.09& 205.83   & 32.69 / 0.8583 & 33.15 / 0.8312 & 38.89 / 0.9320\\
RRDB\cite{wang2018esrgan} &  & 26.93&  114   & 32.76  / 0.8594 & 33.17 / 0.8316  & 38.89 / 0.9322 \\
PSRGAN\cite{huang2021infrared} &  & 0.35&   6.50  & 32.19  / 0.8379 & 32.74 / 0.8136 & 38.27 / 0.9229\\
NLSN\cite{52mei2021image} &  & {44.16}& 210.20    & 32.76  / 0.8589 & 33.24 / 0.8323 & 39.01 / 0.9329 \\
SwinIR\cite{liang2021swinir} &  & 11.90&  53.83   & {32.78}  / {0.8602} &  33.28 / 0.8337 & 39.01 / 0.9329\\
HAT\cite{chen2023activating} &  & 9.62&   54.89  & 32.83  / 0.8595 & 33.32 / 0.8343 & 39.04 / 0.9334 \\
\textbf{LKFormer} &  & {6.52}&  28.31  & \textbf{32.95} / \textbf{0.8616} & \textbf{33.36} / \textbf{0.8348}  & \textbf{39.10} / \textbf{0.9339}\\

\bottomrule
\end{tabular*}
\end{minipage}
\end{center}
\end{table*}

\subsection{Large kernel residual attention}\label{LKRA}
The Transformer architecture has exhibited remarkable advancement in image restoration, owing to its potent representation abilities. Nevertheless, current Transformer-based techniques insufficiently account for the local context in image restoration. Ideally, image restoration methods should be insensitive to shifts in degradation, ensuring the removal of undesirable effects independent of their spatial distribution across the image. Hence, maximizing the information extracted from local regions is indispensable in restoring clean images. Furthermore, the standard Transformer architecture, featuring a self-attention layer, exhibits a quadratic surge in time and memory complexity concerning key-query dot product interactions with input spatial resolution. To circumvent these problems, we present the linearly complex LKRA as an alternative to the standard multi-head self-attention (MSA) module. Assuming that the input image has a size of $h \times w$, the computational complexity of the global MSA module and the LKRA module are as follows:
\begin{equation}
\begin{array}{l l }
    \Omega(\mathrm{MSA})=4 h w C^{2}+2(h w)^{2} C ,
    \\[2mm]
    \Omega(\mathrm{LKRA})=\sum_{i=1}^{n} C\left(K_{i}^{2} +C\right)\times hw
\end{array}
\end{equation}
where C means channel number and K is kernel size. The computational complexity of MSA algorithm escalates quadratically in relation to the size of the image during the feature extraction process. This implies that the processing time commanded by MSA expands exponentially as the image size increases, thereby potentially compromising its efficiency and scalability. In contrast, the computational complexity associated with the application of the LKRA for feature extraction demonstrates a linear scaling pattern with respect to the image size. Consequently, this evidences that LKRA provides enhanced performance and practicality when dealing with images of higher resolution and detail.

As shown in Fig. \ref{fig1:LKFormer} {\color{blue}a}, LKRA incorporates several residual depth-wise convolutional blocks that progressively expand the receptive field during feature extraction, facilitating not only long-range dependency modeling but also local content enhancement. To generate more detailed and high-quality image textures and attain superior visual outcomes, we have integrated the residual concatenation operation multiple times within LKRA. This operation ensures the retention of as many features as possible across the entire hierarchy. Infrared images are inherently susceptible to environmental factors, which often results in them containing less usable information. Therefore, during the SR process of infrared images, it becomes imperative to maximize the utilization of the available input feature information and minimize any potential loss of information during the feature transmission process. Additionally, the incorporation of residual connections can significantly aid network convergence during training and simplify the process of network optimization. In contrast to the standard MSA module that involves acquiring the query and key feature followed by employing the softmax function for attention map computation. To capture the relationships among information from various locations in the feature map, LKRA utilizes a depth-wise convolution with progressively expanding convolution kernels to establish correlations and generate attention map. This approach mitigates the issue of significantly increased complexity resulting from the processing of high-resolution images.

Given a normalized tensor $\mathcal{H}$ of layer, we first encode the cross-channel information and local information by applying a $1\times 1$ convolution followed by a pair of $7 \times 1$ and $1 \times 7$ depth-wise convolution. Subsequently, large kernel residual depth-wise convolution blocks are utilized for capturing long-range dependencies, in order to generate attention maps at a global level. Mathematically, our LKRA can be written as:
\begin{equation}
    f_{k \times k}^{rdwc}\left( \mathcal{H}\right)= \sigma \left(\mathcal{H} + f_{1 \times k}^{d w c}\left( f_{k \times 1}^{d w c}\left( \mathcal{H}\right)\right)\right), 
\end{equation}

\vspace{-8mm}

\begin{equation}
\begin{array}{l l l}
\mathcal{H}_{l_{1}}=f_{1 \times 7}^{d w c}\left(f_{7 \times 1}^{d w c}\left(\sigma\left(f_{1 \times 1}^{c}\left(\mathcal{H}\right)\right)\right)\right), 
\\[2mm]
% \vspace{2mm}
\mathcal{H}_{l_{2}}=f_{31 \times 31}^{r d w c}\left(f_{21 \times 21}^{rd w c}\left(f_{11 \times 11}^{rd w c}\left(\mathcal{H}_{l_{1}}\right)\right)\right), 
\\[2mm]
% \vspace{2mm}
\mathcal{H}_{l_{3}}= f_{1 \times 1}^{c}\left(\mathcal{H} \right) \otimes \mathcal{H}_{l_{2}}, 
\\[2mm]

\mathcal{H}_{l_{out}}= f_{1 \times 1}^{c}\left(\mathcal{H}_{l_{3}} \right),
\end{array}
\end{equation}
where $\sigma$ is SiLU activation function, $f_{k_{1} \times k_{2}}^{dwc}$ represents a depth-wise convolution with the kernel size of $k_{1}\times k_{2}$. To reduce the computational complexity involved in implementing large kernel convolutions, we employ two depth-wise convolution operations to approximate large-kernel depth-wise convolutions within the residual depth-separable convolution block. As an illustration, to mimic a standard 2D convolution with an $11 \times 11$ kernel size, it only requires two convolutions: a $11 \times 1$ and a $1 \times 11$. Here, the kernel size for each residual depth-wise convolution block is set to 11, 21, and 31, respectively.

\begin{figure*}[htbp]
% \footnotesize \resizebox{\textwidth}{60mm}
\centering
\resizebox{\textwidth}{!}{
\begin{tabular}{cccc}
% this is the first column
\hspace{-2mm}
\begin{adjustbox}{valign=t}
    \begin{tabular}{c}
    \specialrule{0em}{0em}{0pt}
    \includegraphics[width=0.35\textwidth]{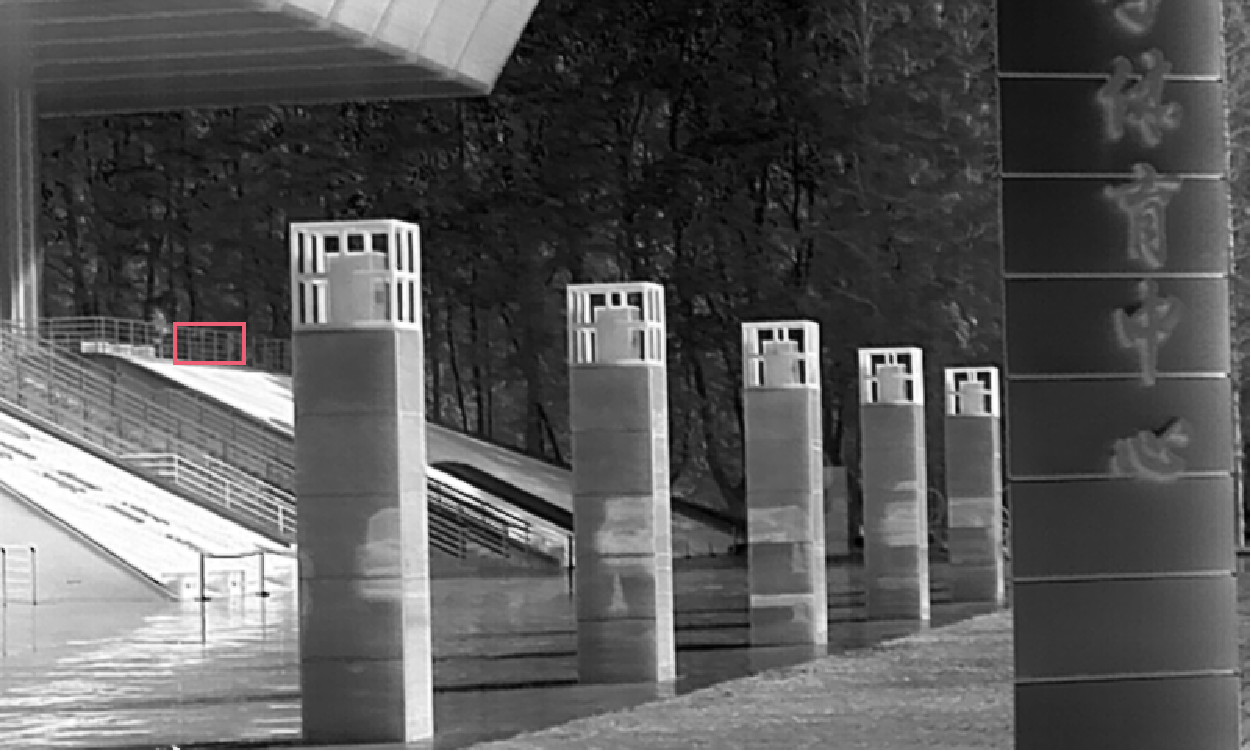}\\ 
    \specialrule{0em}{2pt}{0pt}
    IR700: 3 ($\times$2) \\
    
    \specialrule{0em}{0pt}{7pt}
    \includegraphics[width=0.35\textwidth]{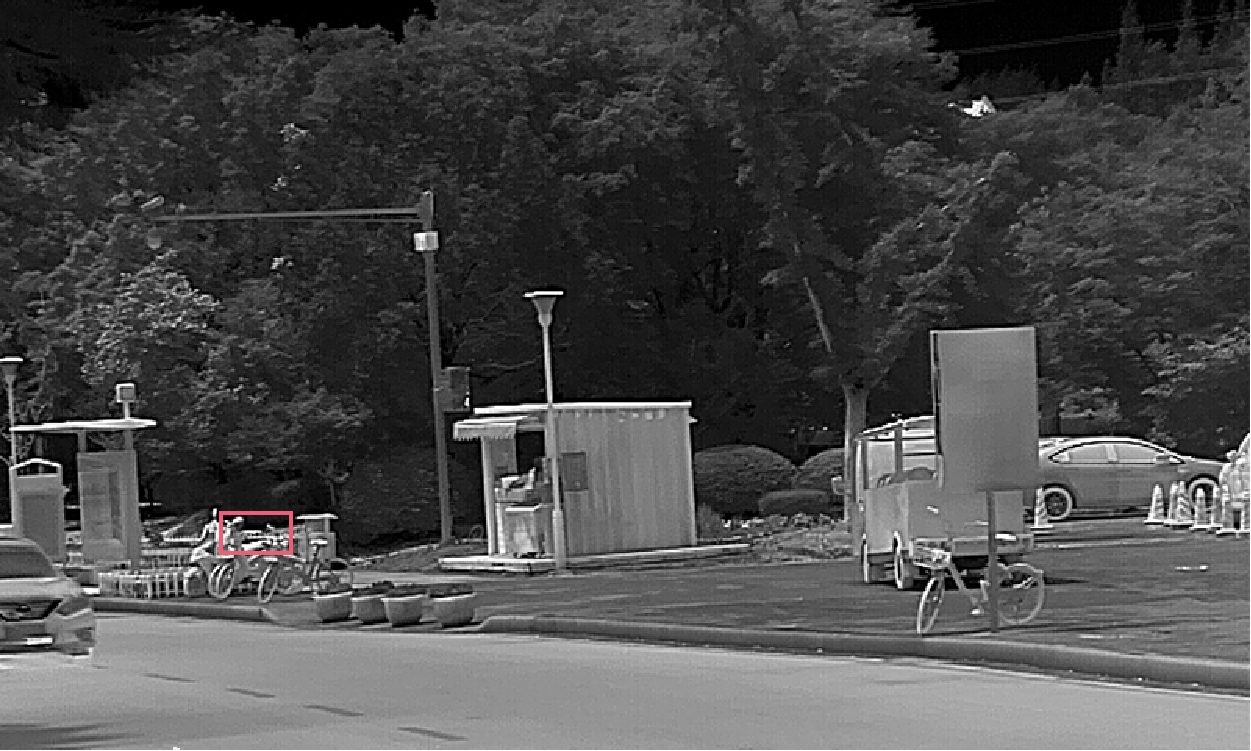} \\
    \specialrule{0em}{2pt}{0pt}
    IR700: 196 ($\times$2)\\
%%%%%%%%%%%%%%%%%%%%%%%%%%%%%%%%%%%%%%%%%%%%%%%%%%%%%%%%%%%%
    \specialrule{0em}{0pt}{10pt}
    \includegraphics[width=0.35\textwidth]{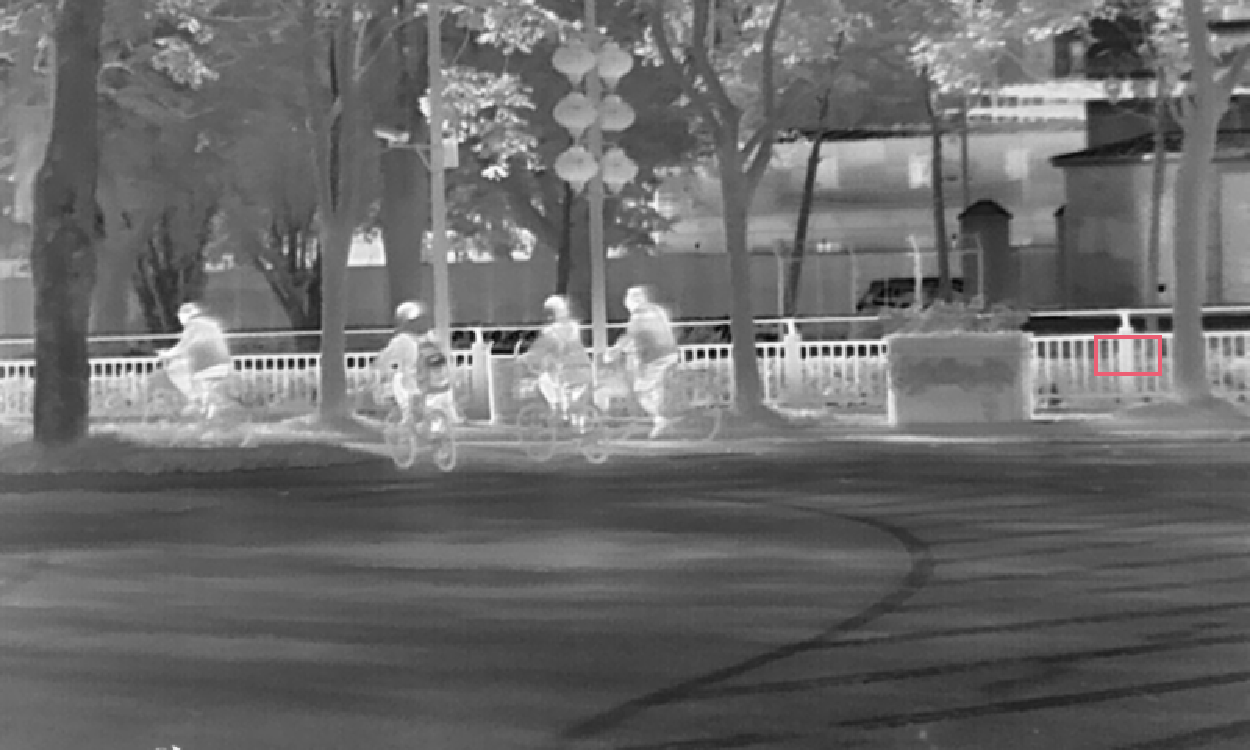} \\
    \specialrule{0em}{2pt}{0pt}
    IR700: 10 ($\times$4)\\
    
    \specialrule{0em}{0pt}{7pt}
    \includegraphics[width=0.35\textwidth]{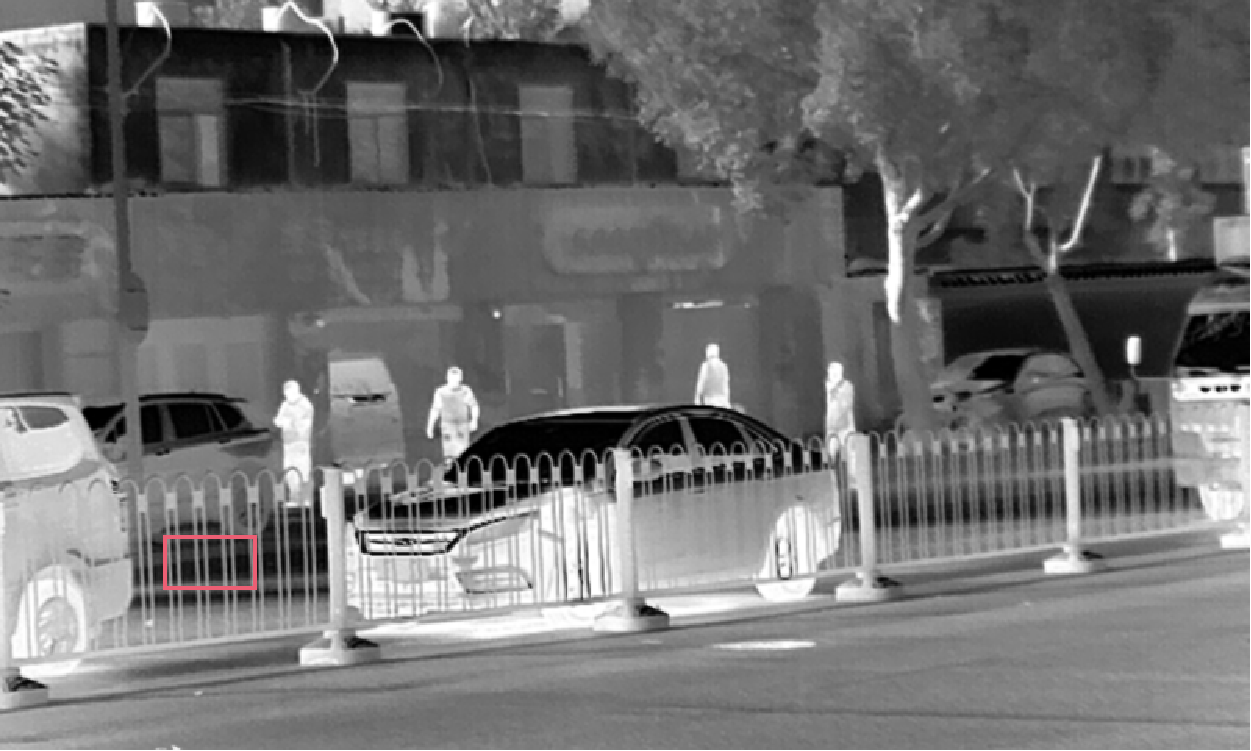} \\
    \specialrule{0em}{2pt}{0pt}
    IR700: 12 ($\times$4)\\
    \end{tabular}
\end{adjustbox}
\hspace{-4mm} 
\begin{adjustbox}{valign=t}
\begin{tabular}{cccc}
\specialrule{0em}{0pt}{0cm}
\includegraphics[width=0.15\textwidth,height=1.25cm]{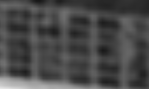}\hspace{-3mm} & 
\includegraphics[width=0.15\textwidth,height=1.25cm]{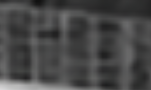}\hspace{-3mm} & 
\includegraphics[width=0.15\textwidth,height=1.25cm]{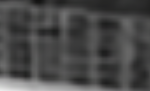}\hspace{-3mm} & \includegraphics[width=0.15\textwidth,height=1.25cm]{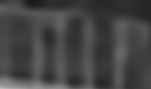}\hspace{-1mm}\\
\specialrule{0em}{-3pt}{0pt}
HR\hspace{-3mm} & EDSR \hspace{-3mm} & RRDB \hspace{-3mm} & PSRGAN \hspace{-3mm}\\
\specialrule{0em}{-3pt}{0pt}
PSNR/SSIM\hspace{-3mm} & 40.94/0.9878 \hspace{-3mm} & 40.50/0.9875 \hspace{-3mm} & 40.37/0.9679 \hspace{-3mm}\\
%%% 
\includegraphics[width=0.15\textwidth,height=1.25cm]{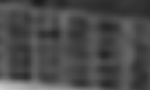}\hspace{-3mm} & 
\includegraphics[width=0.15\textwidth,height=1.25cm]{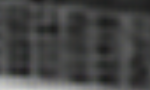}\hspace{-3mm} & \includegraphics[width=0.15\textwidth,height=1.25cm]{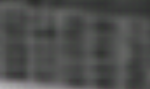}\hspace{-3mm} & \includegraphics[width=0.15\textwidth,height=1.25cm]{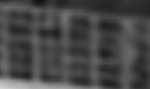}\hspace{0mm} \\
\specialrule{0em}{-3pt}{0pt}
NLSN \hspace{-3mm} & SwinIR \hspace{-3mm} & HAT \hspace{-3mm} & LKFormer\hspace{-3mm}\\
\specialrule{0em}{-3pt}{0pt}
40.87/0.9875\hspace{-3mm} & 40.91/0.9876\hspace{-3mm} & 40.75/0.9873\hspace{-3mm} & \textbf{41.17/0.9886}\hspace{-3mm}\\
\specialrule{0em}{0pt}{7pt} %% 上 下
%%%%%%%%%%%%%%%%%%%%%%%%%%%%%%%%%%%%%%%%%%%%%%%%%%%%%%%%%%%%%%%%%%%%%%%%%%%%%%%%%%%%%%%%%%%%%%%%%%%%%%%%%
\includegraphics[width=0.15\textwidth,height=1.25cm]{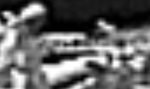}\hspace{-3mm} & 
\includegraphics[width=0.15\textwidth,height=1.25cm]{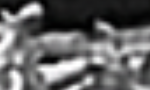}\hspace{-3mm} & 
\includegraphics[width=0.15\textwidth,height=1.25cm]{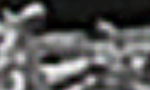}\hspace{-3mm} & 
\includegraphics[width=0.15\textwidth,height=1.25cm]{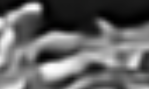}\hspace{-1mm} \\
\specialrule{0em}{-3pt}{0pt}
HR\hspace{-3mm} & EDSR  \hspace{-3mm} & RRDB \hspace{-3mm} & PSRGAN \hspace{-3mm} \\
\specialrule{0em}{-3pt}{0pt}
PSNR/SSIM\hspace{-3mm} & 31.04/0.9065 \hspace{-3mm} & 30.79/0.9060 \hspace{-3mm} & 30.67/0.9048 \hspace{-3mm}\\
\includegraphics[width=0.15\textwidth,height=1.25cm]{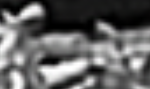}\hspace{-3mm} & 
\includegraphics[width=0.15\textwidth,height=1.25cm]{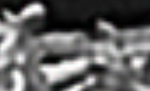}\hspace{-3mm} & \includegraphics[width=0.15\textwidth,height=1.25cm]{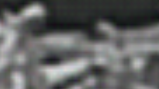}\hspace{-3mm} & \includegraphics[width=0.15\textwidth,height=1.25cm]{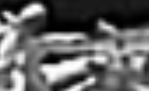}\hspace{0mm} \\
\specialrule{0em}{-3pt}{0pt}
NLSN \hspace{-3mm} & SwinIR \hspace{-3mm} & HAT \hspace{-3mm} & LKFormer\hspace{-3mm}\\
\specialrule{0em}{-3pt}{0pt}
31.16/0.9091\hspace{-3mm} & 31.20/0.9096\hspace{-3mm} & 30.68/0.9045\hspace{-3mm} & \textbf{31.40/0.9110}\hspace{-3mm}\\
%%%%%%%%% x4
\specialrule{0em}{0pt}{7pt}
\includegraphics[width=0.15\textwidth,height=1.25cm]{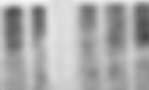}\hspace{-3mm} & 
\includegraphics[width=0.15\textwidth,height=1.25cm]{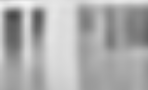}\hspace{-3mm} & 
\includegraphics[width=0.15\textwidth,height=1.25cm]{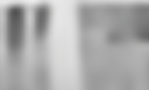}\hspace{-3mm} &
\includegraphics[width=0.15\textwidth,height=1.25cm]{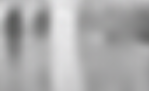}\hspace{-3mm}\\
\specialrule{0em}{-3pt}{0pt}
HR\hspace{-3mm} & EDSR \hspace{-3mm} & RRDB \hspace{-3mm} & PSRGAN \hspace{-3mm} \\
\specialrule{0em}{-3pt}{0pt}
PSNR/SSIM\hspace{-3mm} & 31.32/0.9051 \hspace{-3mm} & 31.65/0.9065 \hspace{-3mm} & 31.19/0.9051 \hspace{-3mm}\\
\includegraphics[width=0.15\textwidth,height=1.25cm]{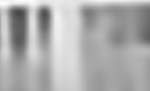}\hspace{-3mm} & 
\includegraphics[width=0.15\textwidth,height=1.25cm]{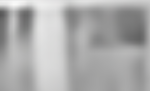}\hspace{-3mm} & \includegraphics[width=0.15\textwidth,height=1.25cm]{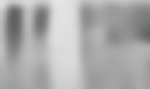}\hspace{-3mm} & \includegraphics[width=0.15\textwidth,height=1.25cm]{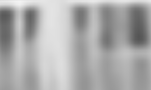}\hspace{-3mm}\\
\specialrule{0em}{-3pt}{0pt}
NLSN \hspace{-3mm} & SwinIR \hspace{-3mm} & HAT \hspace{-3mm} & LKFormer\hspace{-3mm}\\
\specialrule{0em}{-3pt}{0pt}
31.53/0.9070\hspace{-3mm} & 31.59/0.9073\hspace{-3mm} & 31.61/0.9113\hspace{-3mm} & \textbf{31.97/0.9135}\hspace{-3mm}\\
\specialrule{0em}{0pt}{7pt}
%%%%
\includegraphics[width=0.15\textwidth,height=1.25cm]{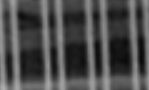}\hspace{-3mm} & 
\includegraphics[width=0.15\textwidth,height=1.25cm]{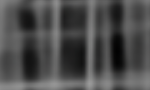}\hspace{-3mm} & 
\includegraphics[width=0.15\textwidth,height=1.25cm]{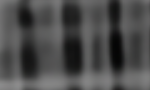}\hspace{-3mm} &
\includegraphics[width=0.15\textwidth,height=1.25cm]{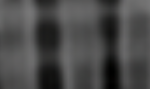}\hspace{-1mm}
\\
\specialrule{0em}{-3pt}{0pt}
HR\hspace{-3mm} & EDSR  \hspace{-3mm} & RRDB  \hspace{-3mm} & PSRGAN  \hspace{-3mm} \\
\specialrule{0em}{-3pt}{0pt}
PSNR/SSIM\hspace{-3mm} & 28.50/0.8843 \hspace{-3mm} & 28.65/0.8876 \hspace{-3mm} & 28.43/0.8839 \hspace{-3mm}\\
\includegraphics[width=0.15\textwidth,height=1.25cm]{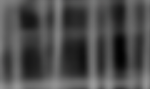}\hspace{-3mm} & 
\includegraphics[width=0.15\textwidth,height=1.25cm]{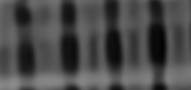}\hspace{-3mm} & 
\includegraphics[width=0.15\textwidth,height=1.25cm]{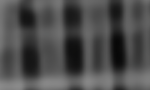}\hspace{-3mm} &  \includegraphics[width=0.15\textwidth,height=1.25cm]{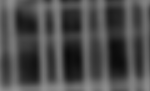}\hspace{0mm}
\\
\specialrule{0em}{-3pt}{0pt}
NLSN \hspace{-3mm} & SwinIR \hspace{-3mm} & HAT \hspace{-3mm} & LKFormer\hspace{-3mm}\\
\specialrule{0em}{-3pt}{0pt}
28.33/0.8832\hspace{-3mm} & 28.94/0.8913\hspace{-3mm} & 29.04/0.8921\hspace{-3mm} & \textbf{29.06/0.8938}\hspace{-3mm}\\

\end{tabular}
\end{adjustbox}
\end{tabular}}
% \vspace{-2mm}
\caption{
Visual comparisons of LKFormer with other SR methods on IR700 dataset. }
\label{fig: visual compare1}
\vspace{-4mm}
\end{figure*} 
%%%%%%%%%%

\subsection{Gated-pixel feed-forward network}\label{subsec4}
Previous researches \cite{dosovitskiy2020image,liu2021swin} have commonly utilized the traditional single-branch FN structure, namely, through the combination of two linear layers or a $1\times1$ convolution followed by a nonlinear activation function situated in between them. Nevertheless, these methods fail to consider the unique demands of dense pixel prediction tasks. To address this issue, we introduce an extra pixel attention branch to complement the regular FN structure, whose primary function is to adjust the weights assigned to each pixel value. The architecture of our gated-pixel FN is shown in Fig.~\ref{fig1:LKFormer} {\color{blue}c}. In fact, it has been previously shown \cite{zamir2022restormer} that controlling the forward flow of features in FN facilitates the model to reconstruct more intricate image textures. Here, in order to minimize the network structure for efficient SR reconstruction of infrared images, we introduce an exclusive pixel attention branch comprising a $1\times1$ convolution for allowing input features to interact among channel dimensions, and a $3\times3$ depth-wise convolution for encoding information from adjacent spatial pixels, thereby enhancing the learning of local image structure. When the feature $\mathcal{F}$ passes through the attention branch, an attention map $\mathcal{M}$ of the same size as $\mathcal{F}$ is generated, and then $\mathcal{M}$ is a dotted product with the $\mathcal{F}$ after $ 1\times 1$ convolutional encoding, thus controlling the forward flow of $\mathcal{F}$. When given an input tensor $\mathcal{F}$, the feature extraction process for this redesigned structure is as follows:
\begin{equation}
\begin{array}{l l}
\mathcal{F}_{l_{1}}= \sigma \left(f_{1 \times 1}^{c}\left(\mathcal{F} \right)\right), 
\\[2mm]

\mathcal{F}_{l_{2}}= f_{1 \times 1}^{c}\left( \mathcal{F}_{l_{1}}\right) \otimes f_{3 \times 3}^{dwc}\left (f_{1 \times 1}^{c}\left( \mathcal{F}_{l_{1}}\right)\right ), 
\\[2mm]

\mathcal{F}_{l_{out}}= f_{1 \times 1}^{c}\left(\mathcal{F}_{l_{2}} \right),

\end{array}
\end{equation}
where $f_{1 \times 1}^{c}$ indicates that the $1\times 1$ convolution, then $f_{3 \times 3}^{dwc}$ denotes $3\times3$ depth-wise convolution. Overall, $f_{3 \times 3}^{dwc}\left (f_{1 \times 1}^{c}\left( \mathcal{F}_{l_{1}}\right)\right )$ learns the attention map $\mathcal{M}$ and modifies the weights of the relevant features of $\mathcal{F}$.

\section{Experiments}\label{sec4}
\subsection{Datasets and Metrics}\label{subsec6}
In our experiments, we utilize the IR700 dataset \cite{ir700zou2021super}, which comprises 700 infrared images, captured via a telescope operating in thermal imaging mode. Each image within this collection presents a resolution of $480\times800$. The predominant content of these images includes pedestrians, vehicles, infrastructure, and buildings. We randomly split the dataset into training and testing sets at an 8:2 ratio. We employ two additional public datasets, which we refer to as results-A \cite{Aliu2018infrared} and ESPOL FIR \cite{danaci2022survey}, to confirm the efficacy of our proposed approach. The results-A is images from fusing infrared and visible images together with the methods reported in the literature \cite{Aliu2018infrared}. The ESPOL FIR dataset consists of 101 infrared images, each with a resolution of $640\times512$. These images were acquired in a variety of environments, both indoor and outdoor, during daylight hours. The primary subjects featured within these images are individuals, vehicles, and various other objects. To assess the quality of the reconstructed infrared images, we adopt two standard evaluation metrics for SR methods, namely, the peak signal-to-noise ratio (PSNR) and the structural similarity index (SSIM) \cite{wang2004image}. Moreover, to better understand the proposed novel Transformer structure, we introduced the local attribution map (LAM) \cite{gu2021interpreting} for overall model analysis.

% ablation table \Checkmark &\XSolidBrush
\begin{table*}[ht]
% \captionsetup{justification=centering}
\centering
% \begin{minipage}{\textwidth}
\caption{Ablations of LKRA for infrared image SR $(\times4)$ on the IR700 dataset. The abbreviation DWC7 refers to the $7\times7$ depth-wise convolution, followed by the abbreviation RDB referring to residual depth-wise convolution block.}\label{ablation-part}%
% \scalebox{1}{
\begin{tabular}{@{}cccccccc@{}}
\toprule
DWC7 & RDB11 &RDB21 &RDB31 & RDB41 & Residual & Params (M)& PSNR (dB)\\
\midrule
\Checkmark &\XSolidBrush & \XSolidBrush & \XSolidBrush & \XSolidBrush & \Checkmark &4.13 &32.71    \\
\XSolidBrush &\Checkmark & \Checkmark & \Checkmark & \XSolidBrush & \Checkmark &4.48 &32.83    \\
\Checkmark & \XSolidBrush& \Checkmark & \Checkmark & \XSolidBrush & \Checkmark &4.00 &32.82  \\
\Checkmark & \Checkmark & \XSolidBrush & \XSolidBrush &\XSolidBrush &\Checkmark &3.02 &32.83 \\
\Checkmark &\Checkmark & \Checkmark & \XSolidBrush &\XSolidBrush &\Checkmark &3.82 & 32.91 \\
\Checkmark &\Checkmark & \Checkmark & \Checkmark &\XSolidBrush &\XSolidBrush &6.52 & 32.80 \\
\Checkmark &\Checkmark & \Checkmark & \Checkmark &\Checkmark &\Checkmark &7.50 & 32.95 \\
\Checkmark &\Checkmark & \Checkmark & \Checkmark &\XSolidBrush &\Checkmark &6.52 &32.95  \\

\bottomrule
\end{tabular}
% }
% \end{minipage}
% \end{center}
\end{table*}

% FN
\begin{table}[ht]
\centering
% \begin{minipage}{174pt}
\caption{Ablation study on influence of the pixel attention branch.}\label{fn}%
\scalebox{1}{
\begin{tabular}{@{}lcc@{}}
\toprule
FN architecture &Params (M)  & PSNR (dB)  \\
\midrule
FN \cite{vaswani2017attention}  & 4.71   & 32.79  \\
GPFN (Ours)   & 6.52   & 32.95    \\
\bottomrule
\end{tabular}
}
% \end{minipage}
\end{table}

% ablation number of LISHB
\begin{table}[ht]
\centering 
% \begin{center}
% \begin{minipage}{\textwidth}
\caption{Ablation study on influence of the number of TL in each RTB.}\label{ablation-number}%
\scalebox{1}{
\begin{tabular}{@{}lcc@{}}
\toprule
Number of TL & Params (M)  & PSNR (dB) \\
\midrule
% \multicolumn{4}{l}{Role of LISHB blocks (n)}\\
n = 2  & 2.56   & 32.76   \\
n = 4   & 4.54   & 32.91    \\
n = 6 (default) & 6.52   & 32.95   \\
n = 8 & 8.50   & 32.95  \\
\bottomrule
\end{tabular}
}
% \end{minipage}
% \end{center}
\end{table}

% ablation number of stage
\begin{table}[ht]
\centering 
% \begin{center}
% \begin{minipage}{\textwidth}
\caption{Ablation study on influence of the number of RTB, each RTB contains 6 TLs. }\label{ablation-stage}%
\scalebox{1}{
\begin{tabular}{@{}lcc@{}}
\toprule
Number of RTB  & Params (M)  & PSNR (dB) \\
\midrule
% \multicolumn{4}{l}{Role of LISHB blocks (n)}\\
n = 2  & 2.40   & 32.74   \\
n = 4   & 4.46   & 32.89    \\
n = 6 (default)   & 6.38 & 32.95   \\
n = 8 & 8.58   & 32.95   \\
\bottomrule
\end{tabular}
}
% \end{minipage}
% \end{center}
\end{table}

\subsection{Results on Infrared Image SR}\label{subsec7}
\begin{figure*}
\begin{minipage}[htbp]{1.0\linewidth}
    \centering
    \begin{overpic}
    [width=.23\linewidth]{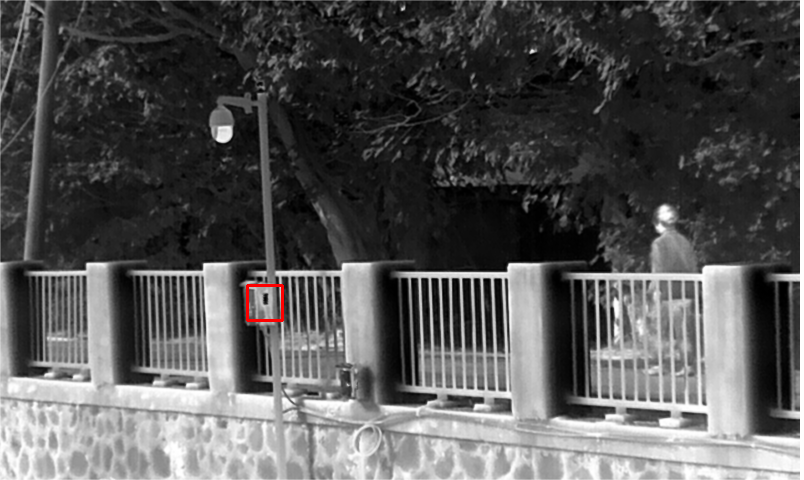} \put(28,-10){\color{black}{\footnotesize IR700: 18}}
    \end{overpic}
    \hspace{1pt} 
    \begin{overpic}
    [width=.23\linewidth]{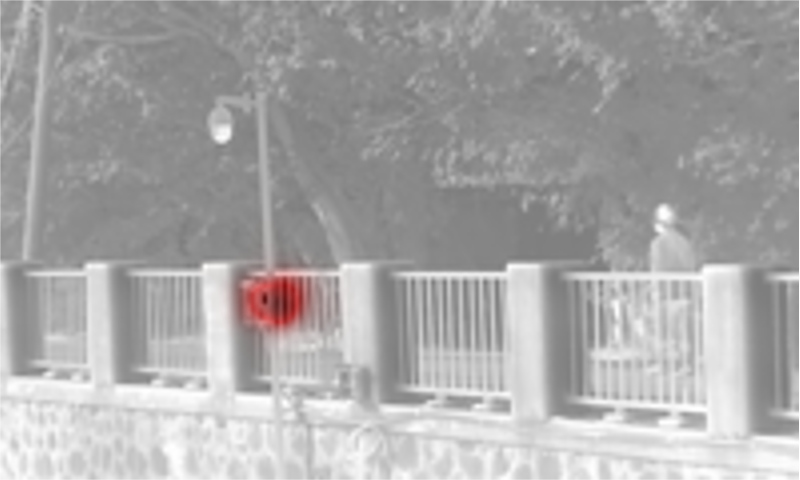} \put(23,-10){\color{black}{\footnotesize SwinIR DI: 0.91}} 
    \end{overpic}
    \hspace{1pt}
    \begin{overpic}
    [width=.23\linewidth]{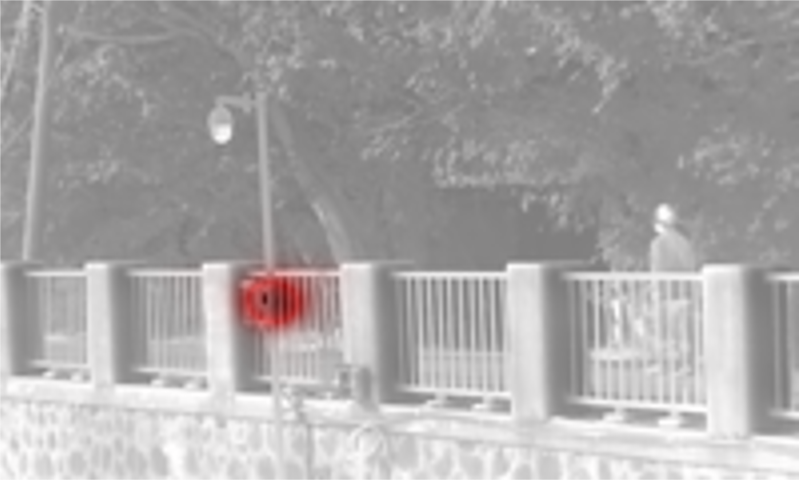} \put(26,-10){\color{black}{\footnotesize HAT DI: 1.29}}
    \end{overpic}
    \hspace{1pt}
    \begin{overpic}
    [width=.23\linewidth]{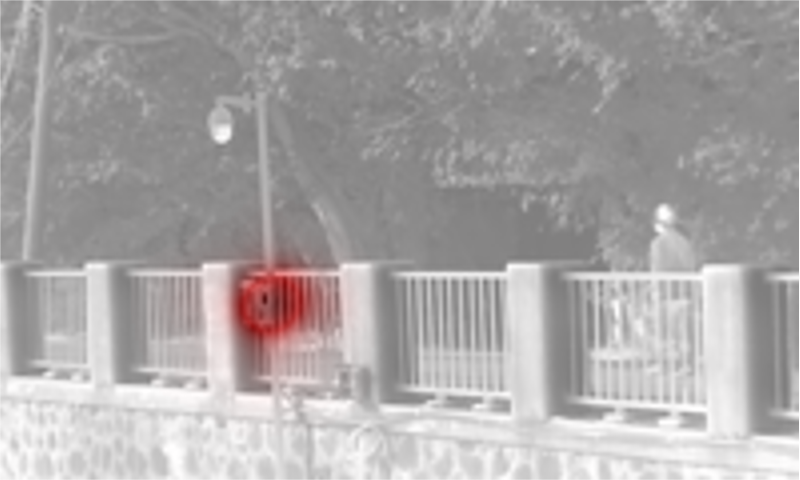} \put(9,-10){\color{black}{\footnotesize LKFormer(Ours) DI: \textbf{2.96} }} 
    \end{overpic}
    \vspace{0.65cm}
    
    \begin{overpic}
    [width=.23\linewidth]{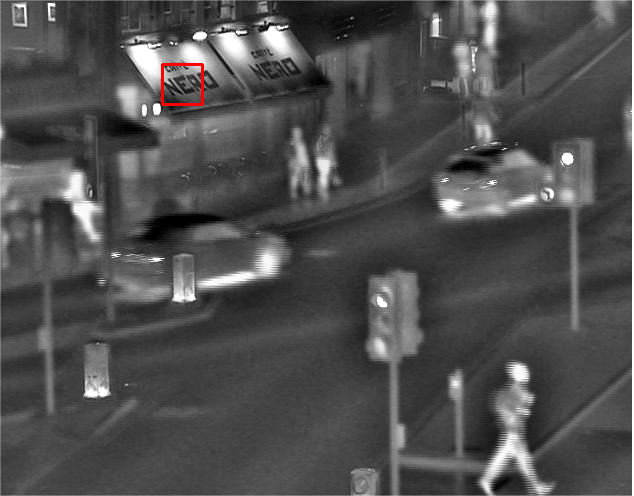} \put(28,-10){\color{black}{\footnotesize result-A: fused2}}
    \end{overpic}
    \hspace{1pt} 
    \begin{overpic}
    [width=.23\linewidth]{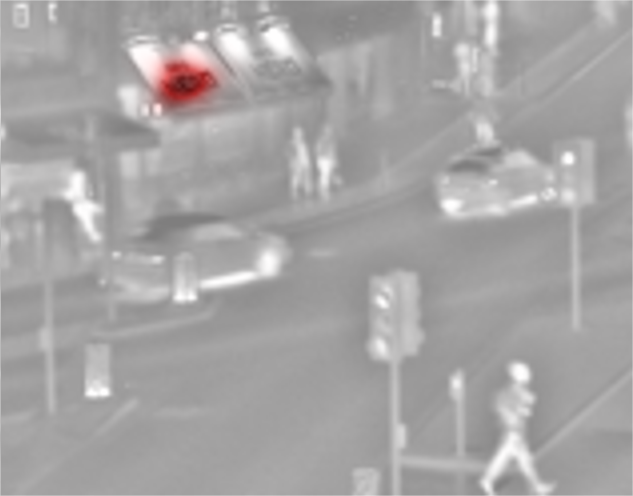} \put(23,-10){\color{black}{\footnotesize SwinIR DI: 0.79}} 
    \end{overpic}
    \hspace{1pt}
    \begin{overpic}
    [width=.23\linewidth]{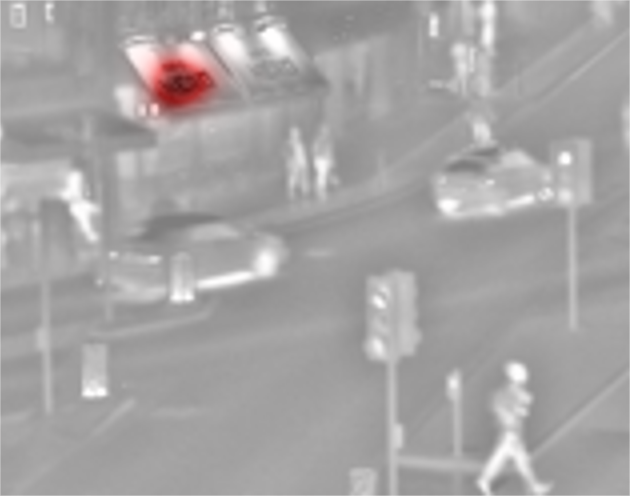} \put(26,-10){\color{black}{\footnotesize HAT DI: 1.18}}
    \end{overpic}
    \hspace{1pt}
    \begin{overpic}
    [width=.23\linewidth]{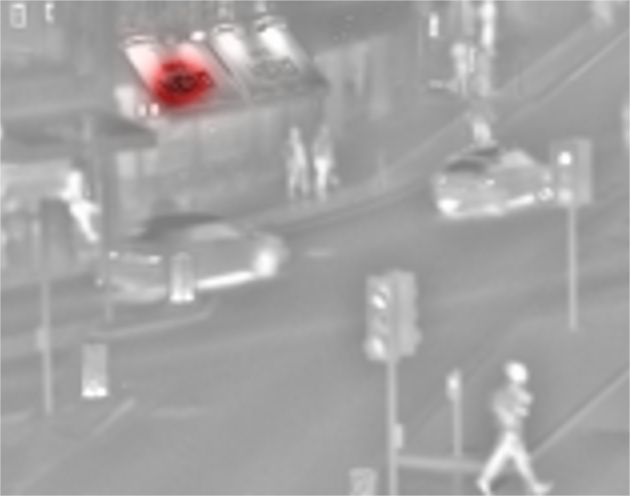} \put(9,-10){\color{black}{\footnotesize LKFormer(Ours) DI: \textbf{1.66} }} 
    \end{overpic}
    \vspace{0.65cm}
    
    \begin{overpic}
    [width=.23\linewidth]{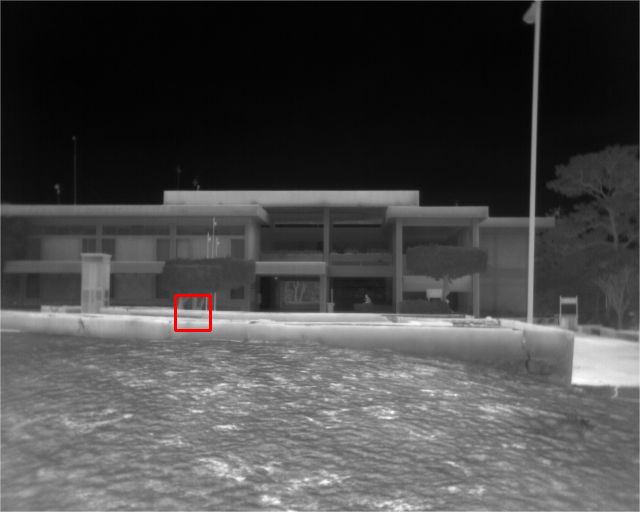} \put(28,-10){\color{black}{\footnotesize ESPOL FIR: 2}}
    \end{overpic}
    \hspace{1pt} 
    \begin{overpic}
    [width=.23\linewidth]{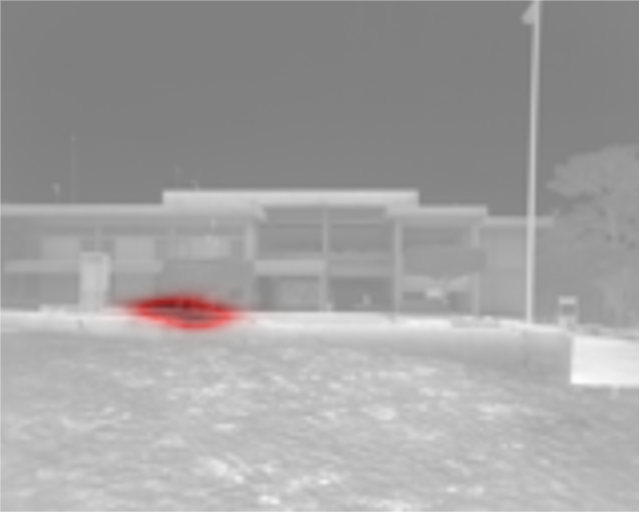} \put(23,-10){\color{black}{\footnotesize SwinIR DI: 0.72}} 
    \end{overpic}
    \hspace{1pt}
    \begin{overpic}
    [width=.23\linewidth]{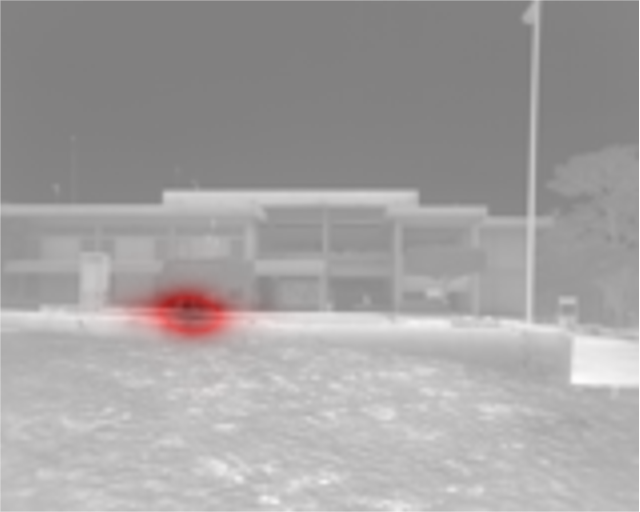} \put(26,-10){\color{black}{\footnotesize HAT DI: 1.61}}
    \end{overpic}
    \hspace{1pt}
    \begin{overpic}
    [width=.23\linewidth]{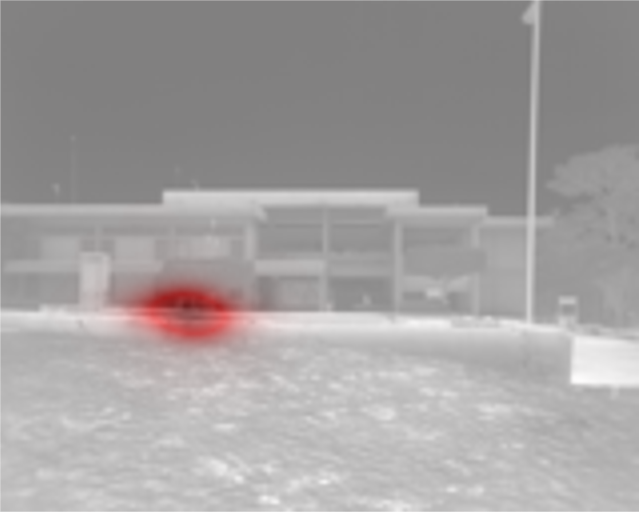} \put(9,-10){\color{black}{\footnotesize LKFormer(Ours) DI: \textbf{2.87} }} 
    \end{overpic}
   
\vspace{0.5cm}
\end{minipage}
\caption{LAM results of SwinIR, HAT, and LKFormer (Ours) in $\times4$ infrared image SR. When reconstructing the patches highlighted in red boxes, the red-marked areas represent informative pixels contributing to the reconstruction. Darker colors indicate a higher degree of contribution. The diffusion index (DI) reflects the range of pixels utilized in image reconstruction, with a higher DI showing a wider range and the involvement of more pixels. The results demonstrate that our method possesses a broader receptive domain, enabling a wider utilization of pixels for image reconstruction.}\label{fig:LAM}
\end{figure*}

\begin{figure*}[htbp]
\centering
\subfigure[PSNR vs. Epochs ($\times2$)]
{
    \begin{minipage}[b]{.45\textwidth}
        \centering
        \begin{overpic}[scale=0.45]{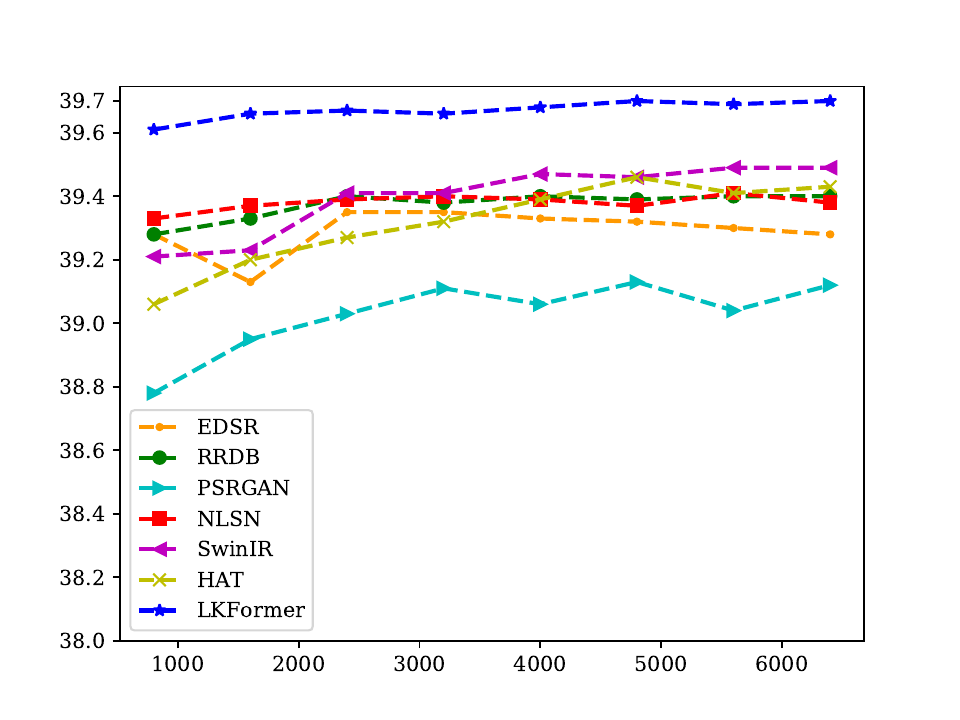}
        \put(25,-0.85){\color{black}{\footnotesize Number of training epochs }}

        \rotatebox{90}{\put(27,-2.5){\color{black}{\footnotesize PSNR (dB) }}}
        \end{overpic}
    \end{minipage}

}
\hspace{4pt}
\subfigure[PSNR vs. Epochs ($\times4$)]
{
    \begin{minipage}[b]{.45\textwidth}
        \centering
        \begin{overpic}[scale=0.45]{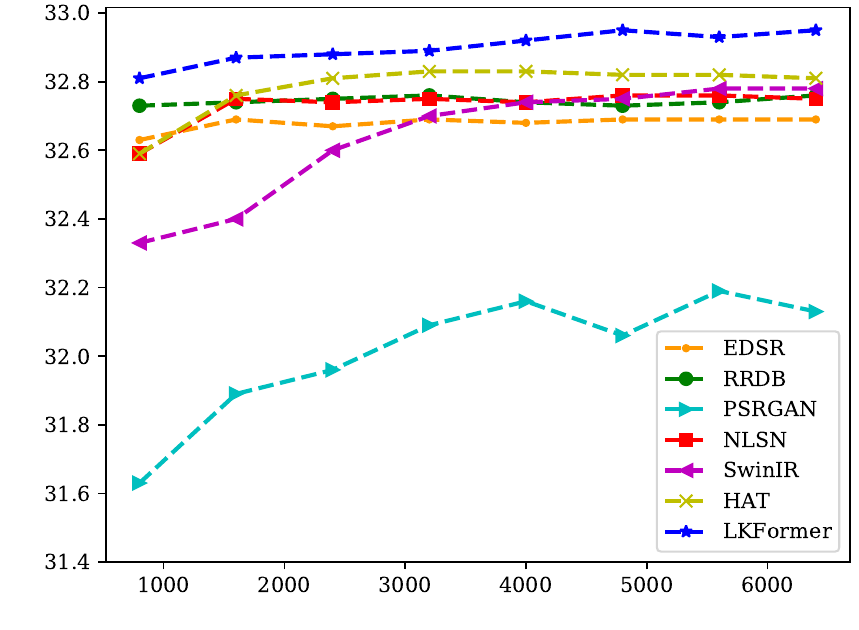}
        \put(25,-0.85){\color{black}{\footnotesize Number of training epochs }}
        \rotatebox{90}{\put(27,-2.5){\color{black}{\footnotesize PSNR (dB) }}}
        \end{overpic}
    \end{minipage}	
}
\caption{The effect of PSNR measure of the different models with respect to training epochs for scaling factors of 2 and 4.}\label{fig:psnr}
\end{figure*}

We compare the proposed LKFormer with state-of-the-art SR methods: EDSR \cite{lim2017enhanced}, RRDB \cite{wang2018esrgan}, PSRGAN \cite{huang2021infrared}, NLSN \cite{52mei2021image}, SwinIR \cite{liang2021swinir}, and HAT \cite{chen2023activating}. Table \ref{tabResult} shows the quantitative comparison results for different upscale factors. Our LKFormer achieves superior SR performance with a reduced parameter count compared to previous state-of-the-art methods. On the IR700 dataset, the maximum gains of PSNR and SSIM access are 0.76dB and 0.0237 for $\times4$, separately. Despite the significantly lower number of parameters in PSRGAN compared to our proposed LKFormer, this reduction also hampers its feature extraction capability, resulting in subpar SR results for infrared images relative to other models. Furthermore, the utilization of a GAN in PSRGAN necessitates the implementation of supplemental tricks to ensure training stability, thereby potentially widening the disparity between PSRGAN and other SR models. In addition, Fig.~\ref{fig:LAM} illustrates the LAM results obtained by various models employing distinct mechanisms. Fig.~\ref{fig:psnr} displays the PSNR results of various models trained on the IR700 dataset, plotted against the number of training epochs. 

Fig.~\ref{fig: visual compare1} shows a visual comparison between LKFormer and the other competitor methods on $\times2$ and $\times4$. The original images are selected from the test dataset of IR700. As we can observe from the zoomed-in view, the infrared images reconstructed by LKFormer are more closely aligned to the ground truth than their competitors regarding edge and texture information. Specifically, Fig.~\ref{fig: visual compare1} demonstrates that LKFormer excels in reconstructing distant objects compared to other methods in the SR reconstruction comparison images labeled 3 and 196. Moreover, in the comparison images labeled 10 and 12, LKFormer significantly preserves the structural integrity of the railings. The visual comparison therefore also further confirms the effectiveness of our proposed LKFormer for the SR reconstruction of infrared images.

%%% LAM

% PSNR图

\subsection{Ablation Study}\label{subsec8}
% we conduct ablation experiments on the IR700 dataset to analyze the effectiveness of our proposed LKFormer. The results of complete experiment are shown in Table \ref{ablation-part} and Table \ref{ablation-number}. Next, we describe the impact of each component.

\textbf{Ablation on LKRA design.} The LKRA module is critical for integrating features from both local and no-local sources in the infrared image SR model. Here, we re-evaluate the method of combining different convolutions within the LKRA module and conducted experiments varying the size of the convolution kernel. Our experimental results, shown in Table \ref{ablation-part}, revealed that the lack of a depth-wise convolution kernel of $7\times7$ prevented the model from focusing effectively on local texture information, resulting in degraded SR performance. Similarly, removing depth-wise convolution kernels of $11\times11$ cause significant variation in kernel size within LKRA, which also let to decreased SR performance. Additionally, the removal of $21\times21$ and $31\times31$ depth-wise convolution limits the model's receptive field and results in a loss of SR performance. To further investigate the contribution of reduced information loss to improve SR in infrared images, we remove only the residual connections inside LKRA and find that the lack of residual structure inside the LKRA module resulted in degraded SR performance. Finally, we incorporate a $41\times41$ depth-wise convolution into the original structure. However, this addition doesn't improve the SR results since the receptive field of the LKRA module is already capable of processing the current input infrared image effectively.

\textbf{Gated-pixel for FN.} The FN module is a crucial component of the Transformer architecture, primarily utilized to enhance the model's nonlinear capability to learn more intricate feature representations. To adopt the Transformer architecture for SR tasks, an extra pixel attention branch is integrated into the standard FN module. Table \ref{fn} illustrates that the gating mechanism present in the GPFN module, which regulates the backward flow of pertinent features, provides an improvement of 0.16 dB over the vanilla FN.

\textbf{Impact of the TL number and RTB
number.} In order to investigate how the number of TLs and RTBs affects the final SR performance of the model, we conduct an ablation study in Table \ref{ablation-number} and \ref{ablation-stage}. The ablation experiments show that if the number of TLs and RTBs is too small, the depth and feature representation ability of the network will be compromised. When there are six TLs and RTBs, the model performs optimally in terms of SR for infrared images. Increasing the number of TLs and RTBs beyond six leads to information redundancy within the network. In order to achieve a trade-off between performance and complexity, the final model structure is comprised of six RTBs, with six TLs nested within each RTB.

\section{Conclusion}\label{sec13}
We present a novel Transformer network, called LKFormer, to solve infrared image SR. Specifically, we design a large kernel residual attention structure to replace the vanilla SA layer to achieve local and non-local feature modeling. In addition, the proposed module can process high-resolution infrared images more efficiently and does not exhibit a quadratic increase in computational complexity with increasing image resolution. Furthermore, to enhance the suitability of the proposed Transformer architecture for the task of dense pixel prediction, we develop a novel module, named the GPFN. The GPFN module improves the information flow within the network by incorporating pixel attention branching. Experimental results show that our LKFormer outperforms competing SR methods while maintaining fewer parameters. However, we only present an intuitive structure until now, there are still many possibilities for improving it. For example, a multi-scale architecture could be introduced to handle images at different resolutions, while a multi-branch structure could enable the network to capture various features at multiple levels. These potential improvements should be explored in future research.  

\section*{Funding}
This work was supported by National Key Research and Development Program of China (No. 2023YFE0114900), Aeronautical Science Foundation of China (No. 2022Z0710T5001), GuangDong Basic and Applied Basic Research Foundation (No. 2022A1515110570), Innovation teams of youth innovation in science and technology of high education institutions of Shandong province (No. 2021KJ088), the Open Project Program of the State Key Laboratory of CAD\&CG (No. A2304), Zhejiang University. The authors would like to thank the reviewers in advance for their comments and suggestions.

\section*{Data availability}
The authors confirm that the data supporting the findings of this study are available in a public repository. These data were derived from the following resources available in the public domain (\href{https://figshare.com/s/2121562561211c0a8101}{https://figshare.com/s/2121562561211c0a8101}, \href{https://github.com/rafariva/ThermalDatasets}{https://github.com/rafariva/ThermalDatasets}).

\section*{Ethics declarations}
\subsection*{Source code}
The source code will be available at \href{https://github.com/sad192/large-kernel-Transformer}{https://github.com/sad192/large-kernel-Transformer}.

\subsection*{Conflicts of interest}
The authors declare that they have no conflict of interest.

\backmatter

\bibliography{sn-bibliography}% common bib file


\end{document}